\documentclass{svjour3}                     
\smartqed  
\usepackage{graphicx}
\usepackage{amsfonts}
\usepackage{amsmath}
\journalname{Few-Body Systems (FB20)}
\begin{document}

\title{
 Efimov Physics in small bosonic clusters
\thanks{Presented at the 20th International IUPAP Conference on Few-Body Problems in Physics, 20 - 25 August, 2012, Fukuoka, Japan}
}
\subtitle{
}


\author{M. Gattobigio         \and
        A. Kievsky            \and
        M. Viviani       
}

\institute{ M. Gattobigio \at 
Universit\'e de Nice-Sophia Antipolis, Institut Non-Lin\'eaire de Nice, CNRS,
1361 route des Lucioles, 06560 Valbonne, France  \\
  \email{mario.gattobigio@inln.cnrs.fr}
            \and
A. Kievsky - M. Viviani \at
Istituto Nazionale di Fisica Nucleare,
Largo Pontecorvo 2, 56127 Pisa, Italy 
}

\date{Received: date / Accepted: date}

\maketitle

\begin{abstract}
We study small clusters of bosons, $A=2,3,4,5,6$, characterized by a resonant
interaction.  Firstly, we use a soft-gaussian interaction that reproduces the
values of the dimer binding energy and the atom-atom scattering length obtained
with LM2M2 potential, a widely used $^4$He-$^4$He interaction. We change the
intensity of the potential to explore the clusters' spectra  in different
regions with large positive and large negative values of the two-body
scattering length and we report the clusters' energies on Efimov plot, which
makes the scale invariance explicit.  Secondly, we repeat our calculation
adding a repulsive three-body force to reproduce the trimer binding energy.  In
all the region explored, we have found that these systems present two states,
one deep and one shallow close to the $A-1$ threshold, and scale invariance has been
investigated for these states.  The calculations are
performed by means of Hyperspherical Harmonics basis set.
\end{abstract}

\section{Introduction}
\label{sec:introduction}
Systems composed by few particles having large value of the two-body scattering
length, $a$,
with respect to the natural length, $\ell$, fixed by the inter-particle potential, 
have been the object of an intense investigation both from a theoretical and experimental 
point of view (for recent reviews see 
Refs.~\cite{braaten:2006_physicsreports,greene:2010_phys.today,ferlaino:2010_physics}).
The interest is driven by their universal properties; the behavior of observables 
do not depend on the microscopical characteristics, namely the inter-particle potential,
but only on symmetries. 

In the limit $a/\ell \rightarrow\infty$, known either as resonant $a\rightarrow\infty$
or as scaling $\ell\rightarrow 0$ limit, the few-particle systems display discrete-scale
symmetry as brought out by Efimov in his works in the early 70's~\cite{efimov:1970_phys.lett.b,efimov:1971_sov.j.nucl.phys.}.
The scaling factor is usually written as $\lambda = \text{e}^{\pi/s_0}$, and
for three-identical bosons $s_0=1.00624$ and $\lambda \approx 22.69$.
This symmetry implies that all observables can be written as an universal 
log-period function of the dimensionless variable $a\kappa_*$, where 
$\kappa_*$ is a three-body parameter encoding the high-energy (short distance)
physics, which enters, at leading order, only through this parameter. 
The fact that the observables' behavior is
governed by discrete-scale symmetry is know as Efimov physics.
For $A=3$, these properties have been studied for
large positive and large negative values of the scattering length in the
$(a^{-1},k)$ plane, with $k={\rm sign}(E)[|E|/(\hbar^2/m)]^{1/2}$, constructing
the  Efimov plot~\cite{efimov:1971_sov.j.nucl.phys.}. 
This plot is useful in the identification of discrete-scale symmetry; in fact, 
if one introduce the radial, $H$, and angular, $\xi$  Efimov variables
by $k = H\sin\xi$ and $ a^{-1} = H \cos\xi$, 
the scale invariance reads $H\rightarrow \lambda^{-1} H$, and $\xi\rightarrow\xi$.

In the present work we extend our previous analysis
of the $A=4-6$ bosonic spectrum~\cite{gattobigio:2011_phys.rev.a} to the $(a^{-1},k)$ plane. We have
modified the strength of the LM2M2~\cite{aziz:1991_j.chem.phys.} potential in order to cover the region of
negative values of $a$ up to $a^0_-$, with this value indicating the threshold
of having a three-body system bound.  We have also increased the intensity of
the interaction in order to extend the analysis to positive values of $a$ in
which the universal character of the system starts to be questionable, i.e,
when the ground-state $E^0_3$ approaches the natural energy
$E_\ell=-\hbar^2/m\ell^2$, which delimits the Efimov window. 

We used the LM2M2 potential to fix the two-body
soft-core potential as in discussed in
Refs.~\cite{kievsky:2011_few-bodysyst.,gattobigio:2011_phys.rev.a}; this has been
possible because of the scale separation between the $^4$He-$^4$He scattering
length, $a=189.41$~a.u., and the natural length $\ell=10.2$~a.u., which is the
van der Waals length calculated for the LM2M2
potential~\cite{braaten:2006_physicsreports}.
In the three-body sector, a three-body soft-core potential is required to
reproduce the ground-state-binding energy of the helium trimer given by
the LM2M2 potential. 

We have performed our numerical calculations in systems with $A\ge 3$, by means
of the non-symmetrized hyperspherical harmonic (NSHH) expansion method with the
technique recently developed by the authors in
Refs.~\cite{gattobigio:2009_phys.rev.a,gattobigio:2009_few-bodysyst.,gattobigio:2011_phys.rev.c,gattobigio:2011_j.phys.:conf.ser.}.
In this approach, the authors have used the Hyperspherical Harmonic (HH) basis,
without a previous symmetrization procedure, and on the representation of the
Hamiltonian matrix, as a sum of products of sparse matrices, well suited for a
numerical implementation.  

As a result, we have observed that in all the region explored the $A=4,5,6$
systems present two states, one deep and one shallow close to the $E_{A-1}^0$
threshold.  This analysis confirms, at least in one zone of the Efimov plot,
previous observations that each Efimov state in the $A=3$ system produces two
bound states in the $A=4$ system, and extends this observation to the $A=5,6$
systems.

\section{Potentials}
\label{sec:potentials}
In our calculation we used
$\hbar^2/m=43.281307~\text{(a.u.)}^2\,\text{K}$ as mass parameter.
The LM2M2 interaction has been modified in the following way 
\begin{equation}
  V_\lambda(r)=\lambda \cdot V_{\text{LM2M2}}(r)\,, 
\label{mtbp}
\end{equation}
and we have varied $\lambda$ from $\lambda=0.883$, where 
$a=a^0_-=-43.84$ a.u., up 
to $\lambda=1.1$ corresponding to $a=44.79$ a.u. The unitary limit 
is produced for $\lambda\approx 0.9743$. 

The two-body gaussian (TBG) potential is 
\begin{equation}
V(r)=V_0 \,\, {\rm e}^{-r^2/R_0^2}\,,
\label{eq:twobp}
\end{equation}
with range $R_0=10$~a.u., 
and we have varied the strength $V_0$ in order to reproduce the values of $a$ given
by $V_\lambda(r)$. 
In the three-body sector we need an hypercentral-three-body (H3B) interaction to better describe
the $A=3$ system obtained with the
modified LM2M2 potential
\begin{equation}
W(\rho_{123})=W_0 \,\, {\rm e}^{-\rho^2_{123}/\rho^2_0}\,,
  \label{eq:hyptbf}
\end{equation}
with the strength $W_0$ tuned to reproduce the trimer energy $E_3^0$ obtained using
$V_\lambda(r)$. 
Here $\rho^2_{123}=\frac{2}{3}(r^2_{12}+r^2_{23}+r^2_{31})$ is the hyperradius
of three particles 
and the range of the three-body force $\rho_0=R_0$  

\section{Results}
\label{sec:results}
\begin{figure}
  \begin{center}
  \includegraphics[width=0.75\linewidth]{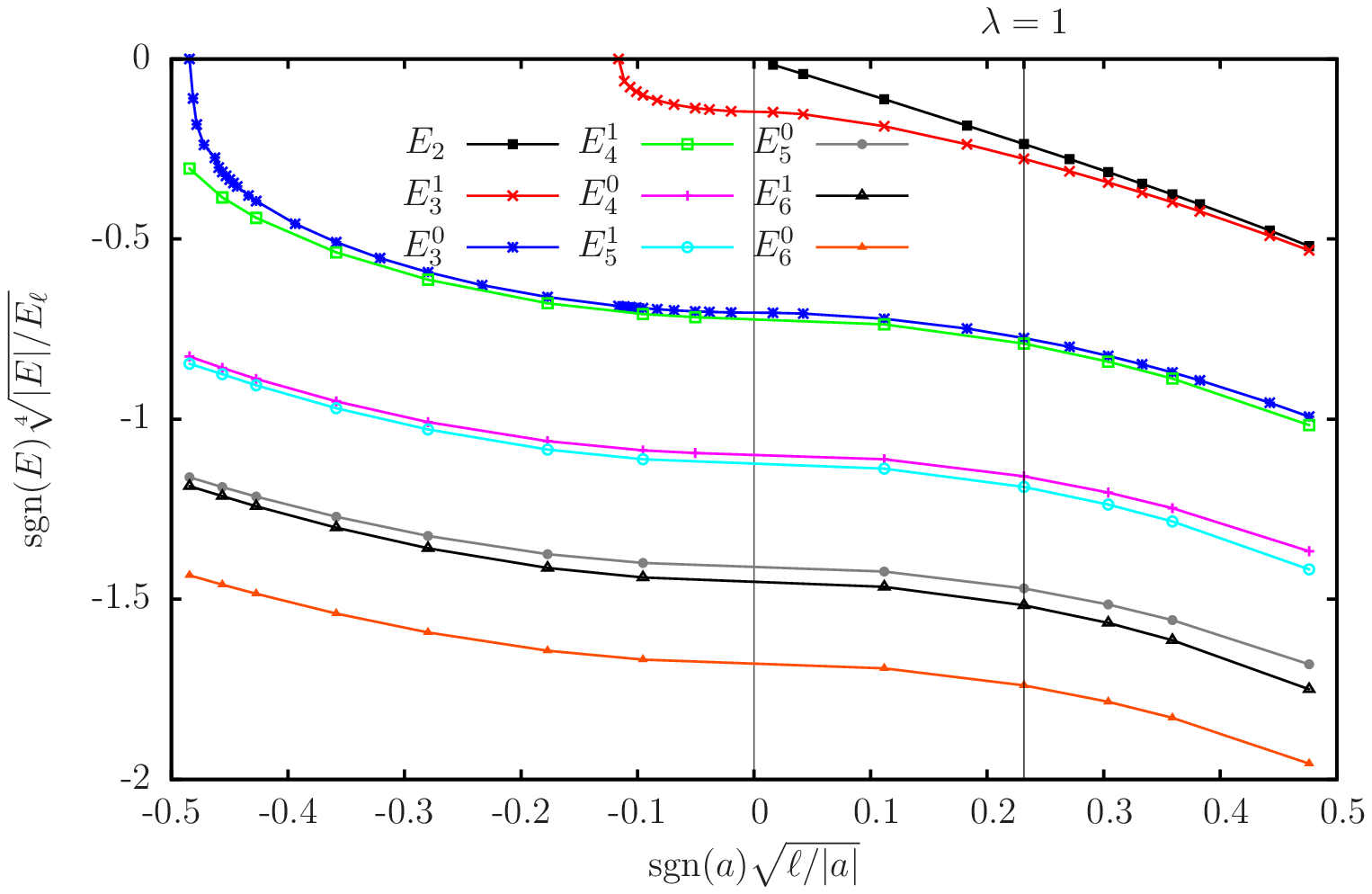}
  \includegraphics[width=0.75\linewidth]{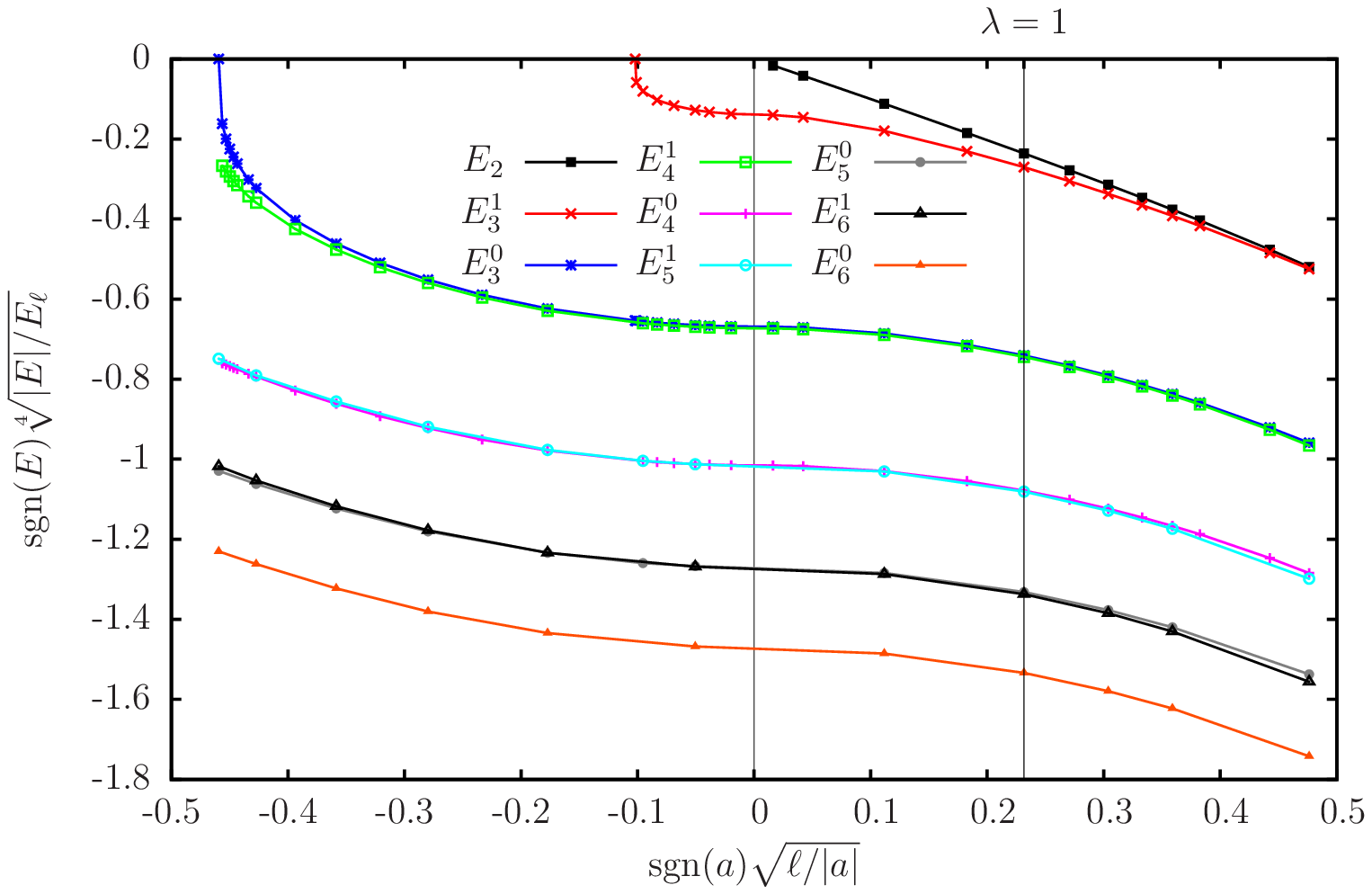}
  \end{center}
  \caption{(color online). Energies of the $A=3-6$ ground and excited 
 states, $E_A^0,E_A^1$, as a function of $a^{-1}$, using
 the two-body gaussian potential (upper panel), and using the two-body plus the 
 hypercentral three-body force (lower panel). In both panels we also give 
 the two-body ground-state energy $E_2$ calculated with the LM2M2 potential.}
 \label{fig:plot2}
\end{figure}
We have solved the $A=3$ problem for bound states using the modified LM2M2
potential, and then we used the resulting energies to fix the strength of the
H3B force. Than we have diagonalized the Hamiltonian for $A=3,4,5,6$ bodies
using the TBG and TBG+H3B potentials. The results are given in
Fig.~\ref{fig:plot2} in two $(a^{-1},k)$ plots, which have been scaled to shrink
the scale factor to $\sqrt{\lambda} \approx 4.8$.

In the case with only the TBG potential, upper panel of Fig.~\ref{fig:plot2},
we observe that the spectrum of the systems $A=4,5,6$ presents two bound
states, one deep and one shallow, for all values of $a$ studied. These calculations
confirms the prediction for $A=4$ of a pair of tetramers attached to a trimer~\cite{platter:2004_phys.rev.a,von_stecher:2009_natphys,deltuva:2011_few-bodysyst.}, and
extends the observation to $A=5$ and $A=6$ systems.

When the
repulsive three-body force is included, lower panel of Fig.~\ref{fig:plot2},
the spectrum moves up and we can
observe that the excited state $E_A^1$ disappears for $A=5,6$ for negative
values of the scattering length as $a$ approaches $a^1_-$. This behaviour
is sensitive to the range of the three-body force $\rho_0$, and it has 
been deeply investigated in Ref.~\cite{gattobigio_2012}.

\begin{figure*}
  \begin{center}
    \includegraphics[width=0.75\textwidth]{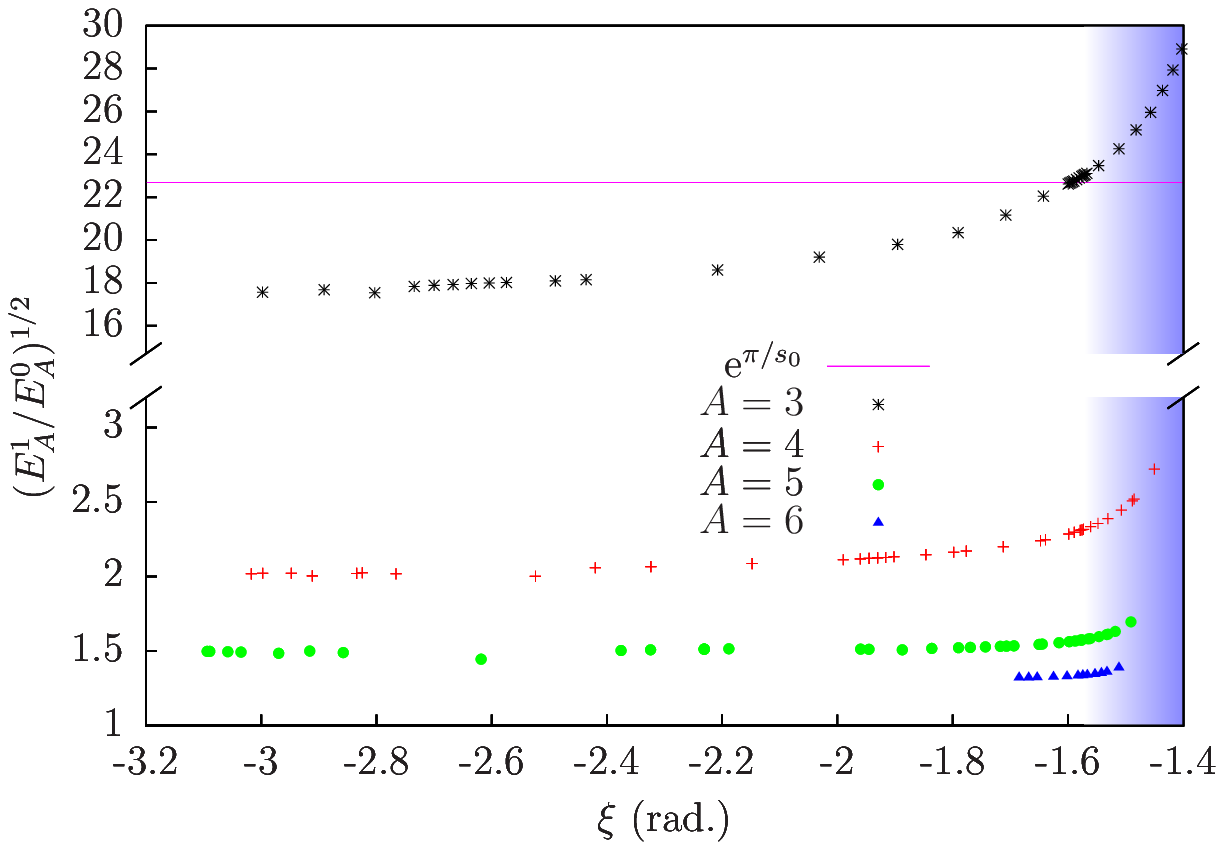}
  \end{center}
\caption{(color online). Square root of the ratios between energies 
of $A$-particle systems as a function of Efimov angle
$\xi$. The shaded region corresponds to positive scattering length.}
\label{fig:rapporti}   
\end{figure*}

In Fig.~\ref{fig:rapporti} we investigate the scale invariance using
Efimov-polar coordinates: for a fixed value of
$\xi$ we report the ratio $(E_A^1/E_A^0)^{1/2}$ calculated with
TBG potential.  For negative values of $a$, the ratios
tend to be constants in agreement with discrete-scale invariance; we note that, 
even if the ratio for $A=3$ tends to be constant, the value is lower than
the expected $\lambda$.
For $a>0$,
which corresponds to the shaded zone in Fig.~\ref{fig:rapporti}, the
non-universal behaviour becomes stronger, and the ratios are no more constants. 

To sum up, we have shown that Efimov physics for $A>3$ manifests with the existence 
of two states for $A=4,5,6$, and that the ratio between each pair tends to be
constant as a function of the Efimov angle $\xi$.


\end{document}